\documentclass[aps,pre,floats,twocolumn,twoside,floatfix]{revtex4-1}

\usepackage{graphicx}
\usepackage{amsmath,amssymb}
\usepackage{psfrag}
\usepackage[utf8]{inputenc}

\begin{document}

\title{Cooperative heterogeneous facilitation:~multiple glassy states
  and glass-glass transition}


\author{Mauro Sellitto}
\affiliation{Department of Information Engineering, Second University
  of Naples, I-81031 Aversa (CE), Italy}

\newcommand{\Tc}{T_{\scriptstyle \rm c}}
\newcommand{\Tg}{T_{\scriptstyle \rm c}}
\newcommand{\TK}{T_{\scriptscriptstyle \rm K}}
\newcommand{\Phic}{\Phi_{\scriptstyle \rm c}}
\newcommand{\pc}{p_{\scriptstyle \rm c}}
\newcommand{\kB}{k_{\scriptscriptstyle \rm B}}

\begin{abstract}
The formal structure of glass singularities in the mode-coupling
theory (MCT) of supercooled liquids dynamics is closely related to
that appearing in the analysis of heterogeneous bootstrap percolation
on Bethe lattices, random graphs and complex networks. Starting from
this observation one can build up microscopic on lattice realizations
of schematic MCT based on cooperative facilitated spin mixtures.  I
discuss a microscopic implementation of the ${\mathsf F}_{13}$
schematic model including multiple glassy states and the glass-glass
transition. Results suggest that our approach is flexible enough to
bridge alternative theoretical descriptions of glassy matter based on
the notions of quenched disorder and dynamic facilitation.
\end{abstract}

\maketitle


Mode-coupling theory (MCT) is considered by many the most
comprehensive first-principle approach to the dynamics of supercooled
liquids~\cite{Gotze}.  Nevertheless, its status is rather problematic
from a fundamental point of view, as the physical nature of the glass
state and the microscopic interpretation of structural arrest are not
yet fully elucidated. This is all the more so when we look at the
higher-order glass singularities in structured and complex liquids.
In this Rapid Communication, I show that multiple glassy states and
glass-glass transition in MCT can be understood in terms of a
generalisation of the notion of dynamic facilitation~\cite{FrAn,RiSo}
and bootstrap percolation~\cite{ChLeRe,BP_rev}. The latter is known to
emerge in a variety of contexts including jamming of granular
materials~\cite{Liu}, NP-hard combinatorial optimization
problems~\cite{MoZe}, neural and immune networks~\cite{Tlusty,
  Stauffer}, and evolutionary modeling~\cite{Klimek}.

The formal structure of glass singularities predicted by MCT is
encoded in the self-consistent equation
\begin{equation}
\label{eq.Phi}
\Phi = (1-\Phi) \ {\mathsf M} (\Phi),
\end{equation}
where $\Phi$ is the asymptotic value of the correlator, and ${\mathsf
  M}$ is the memory kernel describing the retarded friction effect
caused by particle caging, a physical feature associated with the de
Gennes narrowing.  We shall be concerned in the following with
one-component schematic models in which the wavevector dependence of
$\Phi$ is disregarded and ${\mathsf M}$ is a low order polynomial.
Equation~(\ref{eq.Phi})--derived by taking the long-time of the
integro-differential equation describing the evolution of the
correlator of particle density fluctuations--generates a hierarchy of
topologically stable glass singularities, which can be classified in
terms of bifurcations exhibited by the roots of the real polynomial
\begin{equation}
{\mathcal Q}(\Phi) = \Phi - (1-\Phi) \ {\mathsf M} (\Phi) .
\end{equation}
Following Arnol'd notation, adopted in~\cite{Gotze}, an ${\mathsf
  A}_{\ell}$ glass singularity occurs when the corresponding maximum
root of ${\mathcal Q}$ has a degeneracy $\ell$, $\ell \ge 2$, and is
defined by
\begin{equation}
\frac{d^n {\mathcal Q}}{d\Phi^n} = 0 \,, \qquad n=0, \cdots, \ell-1,
\label{eq.dQ}
\end{equation}
while the $\ell$th derivative is nonzero. The polynomial ${\mathcal
  Q}$ has always the trivial root $\Phi=0$, corresponding to a liquid
ergodic state, whereas nonzero values of $\Phi$ correspond to a system
that is unable to fully relax and hence can be identified with a glass
nonergodic state.

For two-parameter systems there are two basic singularities, ${\mathsf
  A}_2$ and ${\mathsf A}_3$, also known as {\em fold} and {\em cusp}
bifurcations. They have been extensively studied by using memory
kernels given by a superposition of linear and nonlinear terms.  In
the ${\mathsf F}_{12}$ schematic model the memory kernel is ${\mathsf
  M}(\Phi) = v_1 \Phi + v_2 \Phi^2$ while the ${\mathsf F}_{13}$ model
is defined by ${\mathsf M}(\Phi) = v_1 \Phi + v_3 \Phi^3$.  The
competition between the two terms produces a variety of nonergodic
behaviors: the linear term gives rise to a continuous liquid-glass
transitions at which $\Phi \sim \epsilon$, where $\epsilon$ is the
distance from the critical point (e.g., $\epsilon=T-\Tc$), while the
nonlinear term induces a discontinuous liquid-glass transition, with
the well known square-root anomaly $\Phi-\Phic \sim
\epsilon^{1/2}$. In the ${\mathsf F}_{12}$ scenario the discontinuous
line joins smoothly the continuous one at a tricritical point. In the
${\mathsf F}_{13}$ scenario, the discontinuous transition line
terminates at an ${\mathsf A}_3$ singularity inside the glass phase
generated by the continuous liquid-glass transition, and therefore
inducing a glass-glass transition (see Fig.~1 for a representative
phase diagram). The scaling form of the order parameter near the
${\mathsf A}_3$ endpoint is $\Phi-\Phic \sim \epsilon^{1/3}$, and more
generally $\Phi-\Phic \sim \epsilon^{1/\ell}$ for an ${\mathsf
  A}_{\ell}$ singularity, as implied by the Taylor expansion of
${\mathcal Q}$ near the critical surface and Eqs.~(\ref{eq.dQ}).  Thus
one can observe a rich variety of nonergodic behaviors whose
complexity is comparable to that of multicritical points in phase
equilibria~\cite{Gilmore}.  It is a nontrivial result that only
bifurcation singularities of type ${\mathsf A}_{\ell}$ can occur in
MCT~\cite{Gotze}.

The ${\mathsf F}_{12}$ and ${\mathsf F}_{13}$ scenarios were first
introduced with the mere intention of demonstrating the existence of
higher-order singularities and glass-glass transition, and then were
subsequently observed in a number of experiments and numerical
simulations of realistic model
systems~\cite{Dawson,Pham,Eckert,Chong,Krako,Kurz,Kim,Sperl,Voigtmann}.
It is important to emphasize that the parameters $v_i$ entering the
memory kernel are smooth functions of the thermodynamic variables,
e.g.  temperature and density, therefore the nature of nonergodic
behaviors predicted by MCT is purely dynamic. This is rather puzzling
from the statistical mechanics perspective of critical phenomena where
diverging relaxation time-scales are closely tied to thermodynamic
singularity.  It has been argued that this unusual situation stems
from uncontrolled approximations.  For example, the intimate
connection of some spin-glass models with MCT has brought to the fore
the existence of a genuine {\em thermodynamic} glass phase at a
Kauzmann temperature $\TK$ below the putative dynamic glass transition
predicted by MCT~\cite{KiWo,KiTh}. A non-trivial Gibbs measure,
induced by a replica-symmetry breaking, would therefore be actually
responsible for the observed glassy behavior~\cite{MePa}.  For this
reason, the nature of the MCT has been much debated since its first
appearance and several approaches have been attempted to clarify its
status~\cite{KiWo,KiTh,MePa,BoCuKuMe,Dave,Andrea,Ikeda,Schmid,Szamel,Silvio}.
I will show here that the idea of dynamic facilitation~\cite{RiSo},
first introduced by Fredrickson and Andersen~\cite{FrAn}, offers some
clues in this direction for its relation with bootstrap percolation
provides a transparent microscopic mechanism of structural
arrest~\cite{ChLeRe,Branco}.  In the dynamic facilitation approach the
coarse-grained structure of a supercooled liquid is represented by an
assembly of higher/lower density mesoscopic \emph{cells}. In the
simplest version a binary spin variable, $s_i=\pm 1$, is assigned to
every cell $i$ depending on its solid or liquid like structure and no
energetic interaction among cells is assumed, ${\mathcal H} = -h
\sum_i s_i$.  The crucial assumption is that the supercooled liquid
dynamics is essentially dominated by the cage effect: fluctuations in
the cells structure occur if and only if there is a certain number,
say $f$, of nearby liquid-like cells. $f$ is called the facilitation
parameter and can take values in the range $0 \le f \le z$, where $z$
is the lattice coordination: cooperative facilitation imposes $f \ge
2$, while non-cooperative dynamics only requires $f=1$.  This very
schematic representation of the cage effect gives rise to a large
variety of remarkable glassy behaviors, and it has long been noticed
that they are surprisingly similar to those found in the dynamic of
mean-field disordered systems~\cite{KuPeSe,Se,SeBiTo}.  It has been
recently observed that in a special case, an exact mapping between
facilitated and disordered models with $\TK=0$,
exists~\cite{FoKrZa}. Since such models are so utterly different in
their premises, it is by no means obvious that such a correspondence
is not accidental and can be extended to systems with higher-order
glass singularities.  To clarify this issue, I will consider a
generalization of the facilitation approach~\cite{SeDeCaAr} in which
every cell $i$ is allowed to have its own facilitation parameter $f_i$
(or, equivalently, an inhomogeneous local lattice connectivity).
Physically, this situation may arise from the coexistence of different
lengthscales in the system, e.g., mixtures of more or less mobile
molecules or polymers with small and large size, (or from a
geometrically disordered environment, e.g., a porous matrix).  In such
facilitated spin mixtures the facilitation strength can be tuned
smoothly and is generally described by the probability distribution
\begin{eqnarray}
  \pi(f_i) & = & \sum_{\zeta=0}^z \ w_{\zeta} \ \delta_{f_i,\zeta} ,
\label{eq.distf.general}
\end{eqnarray}
where the weights $\{w_{\zeta}\}$ controlling the facilitation strength
satisfy the conditions
\begin{eqnarray}
\sum_{\zeta=0}^z w_{\zeta} = 1 ,\qquad 0 \le w_{\zeta} \le 1.
\end{eqnarray}
By tuning the weights one can thus explore a variety of different
situations.  Generally, one observes that when the fraction of spins
with facilitation $f=z-1,z$ is larger than that with $2 \le f \le
z-2$, the glass transition is continuous while in the opposite case it
is discontinuous.  One advantage of the facilitation approach is that
when the lattice topology has a local tree-like structure, one can
compute exactly some key quantities, such as the critical temperature
and the arrested part of correlation and its scaling properties near
criticality. This can be done by exploiting the analogy with bootstrap
percolation.  Let $p$ be the density of up spins in thermal
equilibrium,
\begin{eqnarray}
  p &=& \frac{1}{ 1 + {\rm e}^{-h/\kB T} },
\end{eqnarray}
for a generic spin mixture on a Bethe lattice with branching ratio
$k=z-1$.  As usual, one arranges the lattice as a tree with $k$ branches
going up from each node and one going down, and then proceeds
downwards. In analogy with the heterogeneous bootstrap percolation
problem, the probability $B$ that a cell is in, or can be brought into, the
liquid-like state by only rearranging the state of $k$ cells above
it~\cite{ChLeRe,Branco,SeBiTo,SeDeCaAr}, can be cast in the form
\begin{eqnarray}
  1-B &=& B \ p \ \left\langle \sum_{i=k-f+1}^k {k \choose i}
  B^{k-i-1} (1-B)^i \right\rangle_f,
\label{eq.B}
\end{eqnarray}
where $\left\langle \cdots \right\rangle_f$ represents the average
over the probability distribution Eq.~(\ref{eq.distf.general}). The
right-hand side of Eq.~(\ref{eq.B}) is a polynomial of $1-B$, and
hence the formal structure of Eq.~(\ref{eq.B}) is similar to that of
schematic MCT (once $1-B$ is formally identified with $\Phi$).
Singularities can therefore be classified according to the criteria
already mentioned in the introduction.  Nevertheless, it should be
noticed that what would be the anolog of the MCT kernel in
Eq.~(\ref{eq.B}) can also have negative coefficients (besides
containing an extra term of the form $(1-B)^k/B$), while the
polynomial coefficients of the MCT memory kernel are restricted to
non-negative ones.  In fact, the sets of critical states which specify
some ${\mathsf A}_{\ell}$ glass-transition singularity are not
identical to those describing the full bifurcation scenario of real
polynomials of degree $\ell$, because the coefficients of the
admissible polynomials ${\mathcal Q}$ form only a subset of all real
coefficients. This observation means that the correspondence between
MCT and the heterogeneous facilitation approach is not an identity,
but this still leaves enough room for building up models with MCT
features, although some ingenuity may be required. It has already been
shown, for example, that the ${\mathsf F}_{12}$ scenario is faithfully
reproduced in this framework~\cite{SeDeCaAr,ArSe}. To substantiate the
above observation, I now will focus on the next higher-order glass
singularity, which is the ${\mathsf F}_{13}$ scenario.
\begin{figure} 
\includegraphics[width=8.5cm]{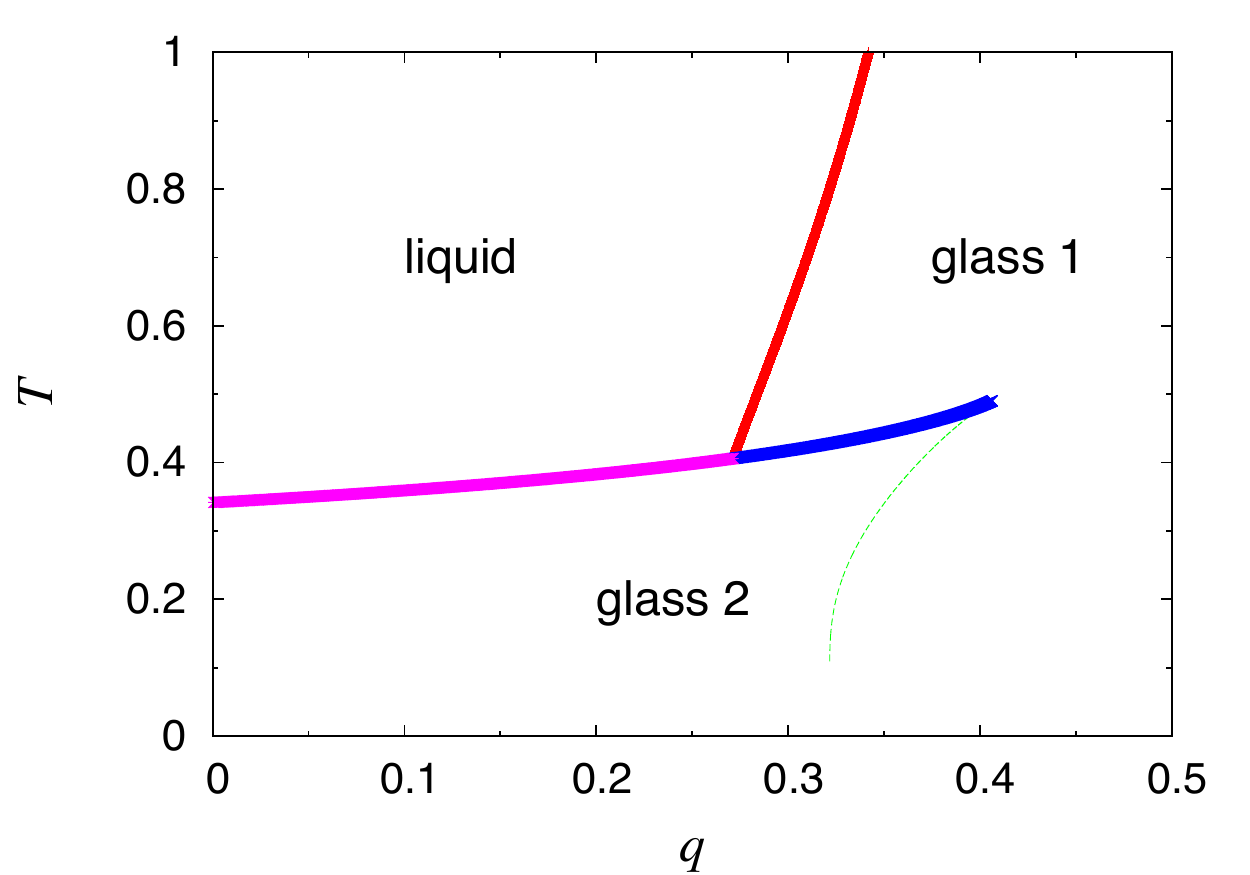}
\caption{Phase diagram for a Bethe lattice with $z=5$ and facilitation
  as in Eq.~(\ref{eq.distf}).  The liquid-glass 1 transition is
  continuous while the liquid-glass 2 and the glass 1-glass 2 are
  discontinuous. The light dashed line is the unstable branch of the
  phase diagram and shows the cuspoid structure of the terminal
  endpoint.}
\label{fig.diag}
\end{figure}
Let us consider, for simplicity, a binary mixture on a Bethe lattice with
$z=5$ and
\begin{equation}
\pi(f_i) = (1-q) \delta_{f_i,2} + q \delta_{f_i,4}.
\label{eq.distf}
\end{equation}
For such a mixture, denoted here as (2,4), the probability $B$ obeys
the fixed-point equation:
\begin{equation}
  1-B = p \left[ q (1-B^4) + (1-q) (1-B)^3 (1+3B)\right].
\label{eq.P245}
\end{equation}
This equation is always satisfied by $1-B=0$, while an additional
solution with $1-B>0$ is found by solving
\begin{equation}
  p^{-1} = 1+B-5B^2+3B^3 + 2 q B^2 (3-B).
\label{eq.P245_1}
\end{equation}
A continuous glass transition is obtained by setting $B=1$ in the
previous equation, giving: $\pc = 1/4q$. Using the relation between $T$
and $p$ (and setting $h/\kB=1$), one gets $\Tc(q) = -1/\ln(4q-1)$,
implying that the continuous transition exists in the range $1/2 \ge q
\ge 1/4$.  The discontinuous transition instead occurs when
Eq.~(\ref{eq.P245_1}) is satisfied and its first derivative with
respect to $B$ vanishes. The latter condition implies
\begin{equation}
  q = \frac{ (9B-1)(1-B)}{6B(2-B)} ,
\label{eq.P245_2}
\end{equation}
and naturally leads to the square-root scaling near the discontinuous
transition line. Thus the discontinuous transition can be graphically
represented by expressing Eqs.~(\ref{eq.P245_1}) and (\ref{eq.P245_2})
in parametric form in terms of $B$. The phase diagram in the plane
$(T,q)$ is shown in the Fig.~\ref{fig.diag}.
It exhibits two crossing glass transition lines, with continuous and
discontinuous nature, corresponding to a degenerate and generic
${\mathsf A}_2$ singularities. The discontinuous branch extends into
the glass region below the continuous line up to a terminal endpoint
which corresponds to an ${\mathsf A}_3$ singularity. The location of
the endpoint is found by simultaneously solving equation
\begin{equation}
  B = \frac{5-6q}{9-6q} ,
\label{eq.P245_3}
\end{equation}
(which is obtained by setting the second derivative of
Eq.~(\ref{eq.P245_1}) to zero), along with Eqs.(\ref{eq.P245_1}) and
(\ref{eq.P245_2}).  The discontinuous branch located between the
crossing point and the endpoint corresponds to a transition between
two distinct glass states, called here glass 1 and glass 2.  They are
respectively characterized by a fractal and compact structure of the
spanning cluster of frozen particles. The passage from one glass to
the other can take place either discontinuously or without meeting any
singularity, i.e. by circling around the endpoint (in a way much
similar to liquid-gas transformation).  The existence of two
transitions in bootstrap percolation was first discovered by Fontes
and Schonmann~\cite{FoSc} in homogeneous trees and then found in
Erd\H{o}s-R\'enyi graphs and complex networks in~\cite{Porto,Cellai}.
However, its relation with glass-glass transition and MCT went
unnoticed. In fact, the correspondence between Eqs.~(\ref{eq.Phi}) and
(\ref{eq.B}) naturally suggests the existence of further singularities
in bootstrap percolation and cooperative facilitated models.

\begin{figure} 
\includegraphics[width=8.5cm]{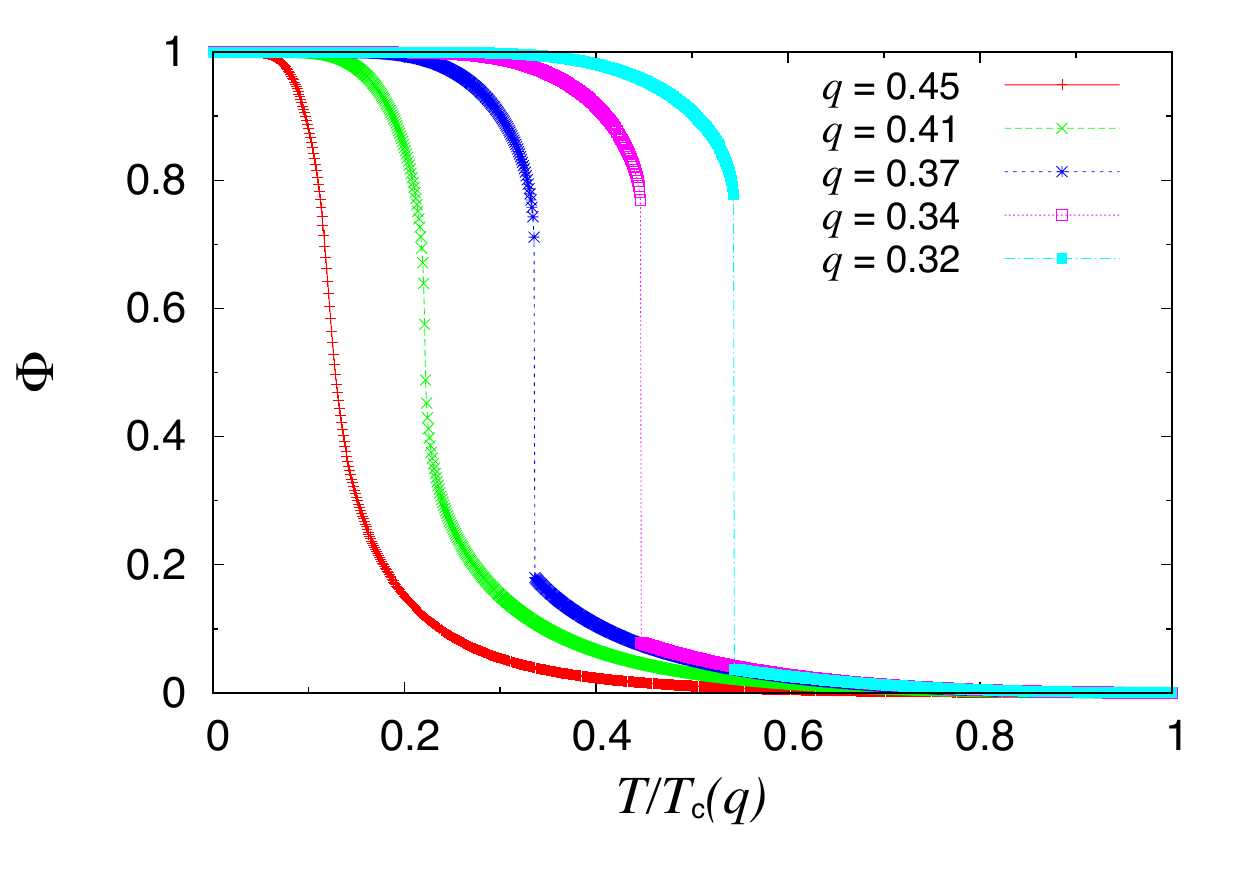}
\caption{ Fraction of permanently frozen spins, $\Phi$, vs temperature
  ratio, $T/\Tc(q)$, with values of $q$ larger than the crossing
  point.}
\label{fig.Phi}
\end{figure}

Fig.~\ref{fig.Phi} reports the behavior of the fraction of frozen
spins, which is the analog of the nonergodicity parameter in the
facilitation approach, when the temperature crosses the liquid-glass
continuous transition and the glass-glass transition.  This quantity
can be exactly computed from $B$~\cite{SeDeCaAr,ArSe}, and its
expression is not reported here--we only notice that its general
features, and in particular the scaling properties near the critical
states, are similar to those of $B$.  We observe that the fraction of
frozen spins first increases smoothly at the liquid-glass continuous
transition and then suddenly jumps at the glass-glass transition. The
jump decreases when $q$ approaches the endpoint and eventually
disappears. At this special point, the additional condition that the
second-order derivative of Eq.~(\ref{eq.P245_1}) with respect to $B$
vanishes, implies a cube-root scaling near the endpoint. These scaling
features are exactly those expected from the ${\mathsf F}_{13}$
scenario, and we obtain similar results for the mixtures (3,5) on a
Bethe lattice with $z=6$.

To summarise, a close relationship exists between the structure
of glass singularities in MCT and that of heterogeneous bootstrap
percolation.  This allows the construction of microscopic realizations
of MCT scenarios based on the heterogeneous cooperative facilitation
approach and provides further insights into the degree of universality
of MCT. The role of the linear and nonlinear terms in the MCT memory
kernel is played in facilitated spin mixtures by the fraction of spins
with facilitation $f=k, k+1$ and $k-1 \ge f\ge 2$, respectively. Their
competition generates continuous and discontinuous liquid-glass
transitions, while the order of singularity is primarily controlled by
the lattice connectivity. This leads to multiple glassy states,
glass-glass transition and more complex glassy behaviors.  In this
framework, the mechanism of structural arrest can be geometrically
interpreted in terms of the formation of a spanning cluster of frozen
spin having fractal or compact structure depending on the continuous
or discontinuous nature of the glass transition.  Finally, from the
relation between MCT and mean-field disordered
systems~\cite{KiWo,KiTh} it follows that quenched disorder and
cooperative facilitation are two complementary, rather than
alternative, descriptions of glassy matter, and this contributes to
the long sought unifying approach to glass physics.

\bibliographystyle{apsrev}

\end{document}